\documentclass[12pt,a4paper,twoside]{article}
\usepackage{amsmath}
\usepackage{amssymb}
\usepackage{amscd}
\usepackage{fleqn}
\usepackage{euscript}
\usepackage{graphics}
\usepackage{pifont}
\usepackage[english]{babel}
\usepackage{latexsym}
\usepackage{epsfig}
\usepackage[small,rm]{caption}
\usepackage{graphicx}
\usepackage{boxedminipage}
\usepackage{chapterbib}
\usepackage{pstricks}
\usepackage{pst-node}
\usepackage{multicol}
\usepackage{subfigure}
\usepackage{amsfonts}
\usepackage{amsthm}
\usepackage{txfonts}

\DeclareFixedFont{\sfracFont}{U}{euf}{b}{n}{7pt}
\newcommand{\prop}[1]{~\guillemotleft\,#1\,\guillemotright~}

\begin{document}
\noindent 
%
%
\begin{centerline}
{\Large{\bf The logic of non-simultaneously decidable propositions}}
\end{centerline}

\begin{centerline}{By {\bf Ernst Specker}, Z\"urich}\end{centerline}
\noindent
\vskip1cm
\noindent
[Translation\footnote{{\scriptsize Completed during a visit at the Center for Time, University of Sydney, Australia. Their hospitality is kindly acknowledged. Thanks are also due to Eric Cavalcanti and R\"udiger Schack for corrections on an earlier version.}} by M.P. Seevinck of  `{\sc  \large die logik nicht gleichzeitig entscheidbarer aussagen}' by Ernst Specker, {\it Dialectica}, vol. 14, 239 -- 246 (1960).]
\vskip1cm
\noindent\begin{flushright}
\emph{La logique est d'abord une science naturelle\footnote{\emph{Logic is in the first place a natural science.}}.}\\
{\sc F. Gonseth}
\end{flushright}

\noindent
\vskip1cm
\noindent
The motto attached to this work is the subtitle of the chapter \emph{La physique de l'objet quelconque}\footnote{\emph{The physics of any object whatsoever}} from the book \emph{Les math\'ematiques et la r\'ealit\'e}\footnote{\emph{Mathematics and reality}}; this physics shows itself to be essentially a form of classical propositional logic, by which, on the one hand, it\footnote{i.e, the classical logic of propositions} obtains a typical realisation, and, on the other hand, it\footnotemark[\value{footnote}] is, in an almost obvious way, deprived of its claim to absoluteness, with which it is occasionally dressed up.  The following remarks concur to this view and can be understood in this same empirical sense.

We assume a domain $B$ of propositions and give ourselves the task of investigating the structure of this domain. Such a structural description of $B$ is only then possible when particular relations or operations are defined between the elements of $B$. The most simple relation may very well be the one of implication ~\guillemotleft\,$a\rightarrow b$\,\guillemotright~ ($a$ and $b$ propositions of $B$), and it will be the basis of the following investigations; we will not assume that the proposition ~\mbox{\guillemotleft\,$a\rightarrow b$\,\guillemotright}~  itself is also a proposition of $B$, when this is not otherwise excluded. Let us for example consider the following example:  The domain of $B$ consists of the ten propositions: ~\guillemotleft\,It is warm\,\guillemotright~, ~\guillemotleft\,It is cold\,\guillemotright~, ~\guillemotleft\,It is raining\,\guillemotright~,~\guillemotleft\,It is snowing\,\guillemotright~,~\guillemotleft\,The sun is shining\,\guillemotright~,
~\guillemotleft\,It is not warm\,\guillemotright~, ~\guillemotleft\,It is not cold\,\guillemotright~, ~\guillemotleft\,It is not raining\,\guillemotright~,~\guillemotleft\,It is not snowing\,\guillemotright~,\prop{The sun is not shining}; the implication \prop{$a\rightarrow b$} holds for some $a,b$ in  $B$, for some pairs it certainly does not hold, whereas for others it may remain doubtful; examples are \prop{If it is warm, then it is not cold},\prop{If it is cold, then it is snowing},\prop{If it is raining, then it is not snowing}. In the following we will not concern ourselves with the problem indicated by the third example, namely that the implication \prop{$a\rightarrow b$} can be doubtful for some pairs: for arbitrary two propositions $a,b$ of $B$ it is the case that either \prop{$a\rightarrow b$} holds, or it does not hold. Let us also call attention to the second example \prop{If it is cold, then it is snowing}. Of this implication we have said that it does not hold. But with this it is of course not asserted that  it can not be cold and snowing, but only that it is not always the case that when it is cold, it is also snowing.  This indicates that the propositions \prop{It is cold}, etc, are not meant as abbreviations for something like \prop{It is cold at 11.50 hrs. am on May 1$^{\mathrm{st}}$ at the garden gate of the property at 60 Goldauer street in Z\"urich} 
(maybe with further included more precise specifications, in case these have been omitted) but in general (as \prop{forms of propositions}), in the way they enter in the formulation of natural laws.

On the basis of the implication it is then possible to indicate when a proposition $c$ of $B$ can be regarded to be the conjunction of the propositions $a,b$ of $B$: For that it is firstly necessary that the implications \prop{$c\rightarrow a$} and \prop{$c\rightarrow b$} hold (if $a$ and $b$, then $a$ ; if $a$ and $b$, then $b$)  and that $c$ obtains the following extremal requirement: if 
for some $c'$ in $B$ it is the case that \prop{$c'\rightarrow a$} and \prop{$c'\rightarrow b$}, then also \prop{$c' \rightarrow c$} (if it is the case that $c'$ implies $a$ and $c'$ implies $b$, then also $c'$ implies $a$ and $b$). It is now by no means self-evident that the domain $B$ contains an element that has these properties; in the example of ten propositions given above there is, for example, no conjunction for any pair of distinct elements. However, it is of course not excluded by this example that to a domain $B$ there exists  a more encompassing domain $B'$ that has this closure, and more than this is not meant when it is said that to two arbitrary propositions one can always associate  a conjunction. However, before we turn to this question it must be investigated whether the conjunction of two propositions is uniquely determined. In case both $c_1$ and $c_2$ are conjunctions of $a$ and $b$, then, according to our assessment, the implications \prop{$c_1\rightarrow c_2$} and \prop{$c_2\rightarrow c_1$} hold (for which we also write  \prop{$c_1\leftrightarrow c_2$}, and we say, $c_1$ and $c_2$ are equivalent). Equivalent propositions need not be identical (Example: \prop{it is lightening and it is thundering}, \prop{it is thundering and it is lightening}); because of this, if uniqueness of the conjunction (and of the other combinations) is desired, one considers instead of the propositions their equivalence classes, and it is shown that the equivalence class of the conjunction of two propositions only depends on the equivalence classes of the related propositions. In the case of classical logic one is directed in this way to the Boolean lattices; however, an analogous procedure is possible in any other calculus of logic that can be considered (such as in the intuitionistic, modal, or multiple-valued logic). The possibility of the transition to equivalence classes assumes in the first place that the proper meaning of the relation \prop{$c\leftrightarrow d$} is an equivalence relation, that is, it has the properties of reflexivity  (\prop{$c \leftrightarrow c$}), symmetry (if \prop{$c \leftrightarrow d$}, then  \prop{$d \leftrightarrow c$}) and of transitivity (if \prop{$c \leftrightarrow d$} and \prop{$d \leftrightarrow e$}, then also \prop{$c \leftrightarrow e$}). Of these properties the one of symmetry is fulfilled because of the definition of \prop{$\leftrightarrow$} out of  the implication \prop{$\rightarrow$}; the reflexivity \prop{$c \leftrightarrow c$} is obtained from the existence of the implication \prop{$c \rightarrow c$}. Because up until now we have not made any assumptions about the implication, evidently \prop{$c\rightarrow c$} can not be proven, and indeed the analysis of the concept of implication that will follow below gives us no reason to start from \prop{$c\rightarrow c$}. The transitivity of the relation \prop{$\leftrightarrow$}  is usually concluded from the transitivity  of the implication: In case \prop{$c\rightarrow d$} and \prop{$d\rightarrow e$}, then also \prop{$c \rightarrow e$}. It 
could subsequently appear as if this transitivity as well as the existence of \prop{$c\rightarrow c$} is so tightly connected to the concept of implication that it would be pointless to call a non-transitive relation an \prop{implication}. That this is not quite true should be exemplified by the following story, which is situated a long time ago and in a far away country.

During the age of king Asarhaddon a wise man from Ninive taught at the school of prophets in Arba'ila.
He was an outstanding representative of his discipline (solar and lunar eclipses) who was, except for the heavenly bodies,  concerned almost exclusively about his daughter. His teaching success was modest, the discipline was proved to be dry and required also previous mathematical knowledge, which was scarcely present. 
Although he did not find the interest amongst the students that he had hoped for, it was however given to him in abundance in a different field. No sooner had his daughter reached the marriageable age, than he was bombarded with marriage proposals to her by students and young graduates. And although he did not believe that he could keep her to himself forever, she was in any case still far too young and her suitors were also in no way worthy of her.  And in order for each of them to convince themselves that they were unworthy, he promised them that she would be the wife of he who would solve a prediction-task that was posed to them. The suitors were led in front of a table on which three boxes were positioned in a row, and they were ordered to indicate which of the boxes contained a gem and which were empty. And now no matter how many times they tried, it seemed to be impossible to solve the task. After their predictions, each of the suitors was ordered to open two boxes which they had indicated to be both empty or both not empty:  it turned out each time that one contained a gem and the other did not, and, to be precise, sometimes the gem was in the first, sometimes in the second of the boxes that were opened. But how can it be possible that from three boxes neither two can be indicated as empty, nor as not empty? The daughter would have remained unmarried until the father's death, if she would not have swiftly opened two boxes herself after the prediction of the son of a prophet, who indicated that precisely one should be filled and the other empty, which turned out to be actually the case.  At the weak objection by the father that he would have opened two other boxes, she attempted to open the third box, which turned out to be impossible, after which the father declared in a mumbling way the not-falsified prediction as valid.

In order to logically analyse the mentioned prediction-task we introduce the following six propositions $A_{i}^{ }$, $A^*_{i}$ $(i=1,2,3)$, where $A_{i}^{ }$ indicated that the $i$-th box is filled,  $A_{i}^{*}$ that it is empty. The attempts made by the suitors  indicate that in the domain of these propositions the following implications hold:  $A_{i}^{ } \rightarrow A_{j}^{*}$, $A_{i}^{*} \rightarrow A_{j}^{ }$ (for each pair $i,j$ of different numbers 1,2,3); of course, also the implications   $A_{i}^{ } \rightarrow A_{i}^{}$, $A_{i}^{*} \rightarrow A_{i}^{*}$  ($i=1,2,3$) hold. Also, the implications  $A_{1}^{} \rightarrow A_{2}^{*}$, $A_{2}^{*} \rightarrow A_{3}^{ }$ hold, whereas   $A_{1}^{ } \rightarrow A_{3}^{ }$ does not hold, but only $A_{1}^{} \rightarrow A_{3}^{*}$.  It is clear that, why not a single one  of these three implications can be refuted, is only because it is impossible to open all three boxes. We have hereby found an assumption without which the deduction from the implication \prop{$a\rightarrow b$}, \prop{$b\rightarrow c$} to the implication \prop{$a\rightarrow c$} is not possible without anything further ado:  All of the propositions $a,b,c$ must be verifiable together. (The implication \prop{$a\rightarrow b$} must of course always be regarded in such a way that $a$ and $b$ are verifiable together and that always when this is performed, with $a$ also $b$ is satisfied.)

The difficulties that arise from propositions that are together not decidable emerge very clearly from propositions about quantum mechanical systems. In accordance to the there commonly used terminology [i.e., in that field], we would like to call the collection of such propositions as not-simultaneously decidable; the logic of quantum mechanics was investigated for the first time by von Neumann and Birkhoff \cite{vNB}. We will return to their results later.
 In a certain sense the scholastic speculations about the \prop{Infuturabilien}\footnote{To be translated as something like `future contingencies'.} also belong here, that is, the question whether the omniscience of God also extends to events that would have occurred  in case something would have happened that did not happen. (cf. e.g. \cite{S}, Vol. 3, p. 363.)

When we consequently take into account that not every collection of propositions is simultaneously decidable, then to the description of the structure of a collection $B$ of propositions belongs, besides the implication, also the set $\Gamma$  of subcollections of $B$ that are simultaneously decidable. If for two elements $a,b$ of $B$ 
it is the case that \prop{$a\rightarrow b$} holds, then $(a,b)$ is in $\Gamma$. Because we assume in particular for each $a$ that \prop{$a\rightarrow a$} $(a)$ is in $\Gamma$, that is, $B$ does not contain any undecidable propositions. We now further assume that the implication is transitive, and, consequently, that under \prop{$\leftrightarrow$} $B$ falls apart in classes of equivalent propositions. However, in order to be able to go from $B$ to the collection $B'$ of equivalence classes, we need the further assumption that the set $\Gamma$ is compatible with the arrangement into classes, that is, for example, when $(a,b)$ is in $\Gamma$ and it is the case that \prop{$a\leftrightarrow a'$}, that then also $(a',b)$ is in $\Gamma$. This we would now like to assume and we then get a collection $B'$ with a relation \prop{$\rightarrow$} which partly orders $B'$, and also a set $\Gamma'$ of subsets of $B'$; $\Gamma'$ contains all singletons, with each set [it contains all] its subsets,  and in case \prop{$a \rightarrow b$} the set $(a,b)$. Birkhoff and von Neumann have shown that the set $B'$ that is assigned in this way to the set of propositions $B$ about a quantum mechanical system is isomorphic to the  collection of linear and closed subspaces of a complex Hilbert space (which in particular cases can be  a unitary space); the implication corresponds to the relation of containment. A collection $C$ of subspaces corresponds precisely then to a collection from $\Gamma'$, when there is a unitary basis for the space which contains a basis for each subspace of $C$. It can be shown that this is already the case when there is two by two such a basis for spaces from $C$; this requirement is precisely then obeyed when, in the sense of elementary geometry, the subspaces are orthogonal, that is, when the completely orthogonal complement of the intersection cuts the subspaces into completely orthogonal spaces. A collection of propositions about a quantum mechanical system is consequently precisely then simultaneously decidable, when they are two by two [simultaneously decidable]. Further, it can easily be shown that each such collection of propositions is contained in a Boolean lattice, that is, for them classical logic holds. (A corresponding assumption appears also in a general theory as natural.) It is in particular  the case that a negation $\neg a$ is related to every proposition $a$; $\neg a$ is only then simultaneously decidable with $b$ when $a$ is [simultaneously decidable] with $b$.  To two simultaneously decidable propositions $a, b$ is related a conjunction and disjunction, and all these propositions are simultaneously decidable. On the basis of the above indicated characterization it is possible to also relate a conjunction and, analogously, a disjunction to non-simultaneously decidable propositions;  in the totality of subspaces of a Hilbert space these operations correspond to the intersection and the spanned subspace. 
In contradistinction to the work of Birkhoff and von Neumann we have to give this up here, because it is essential to the 
problems to be considered that the operations are only defined for simultaneously decidable propositions. For we want to devote ourselves to the question whether it is possible to 
embed the totality of (closed) subspaces of a Hilbert space into a boolean lattice in such a way that the negation and also the conjunction and disjunction, in so far as they are defined (that is, for orthogonal subspaces) retain their meaning. The question can also be formulated more visually in the following way: Is it possible to extend the description of a quantum mechanical system through the introduction of supplementary --- fictitious --- propositions in such a way that in the extended domain the classical propositional logic holds (whereby, of course, for simultaneously decidable proposition negation, conjunction and disjunction must retain their meaning)?
 
The answer to this question is negative, except in the case of Hilbert spaces of dimension 1 and 2. In the case of dimension 1, the lattice of subspaces is the  Boolean lattice of two elements. In the case of dimension 2, the lattice of subspaces can be described in the following way: 
 Consider subspace $H$ (total space), $O$ (space consisting of the null vector) and $A_a$, $B_a$ (with $a$ a set that ranges over the cardinality of the continuum); $H$ and $O$ are orthogonal to all subspaces, also $A_a$ and $B_a$ [are orthogonal] to $H$, $O$, $A_a$ and $B_a$. The complement (negation) of $A_a$ is $B_b$ and vice versa, the negation of $H$ is $O$ and vice versa. The conjunction of $A_a$ and $B_a$ is $O$, their disjunction is $H$; $H$ and $O$ are the identity element  and null element of the \prop{partial lattice}: $O\vee C=C$, $O\wedge C=O$, $H\vee C=H$, $H\wedge C=C$ ($C$ is an arbitrary subspace). It is easy to see that this structure can be embedded in a Boolean lattice. That such an embedding is not possible from dimension 3 and higher follows from the fact that it [this embedding] is not possible for a three-dimensional space.
For the sake of visualisability we will restrict ourselves to the real orthogonal space which is contained in the unitary space, for which the embedding-protocol then is the following: The totality of linear subspaces of a three-dimensional orthogonal vector space can be mapped one-to-one to a Boolean lattice in such a way that for arbitrary orthogonal subspaces $a,b$ the following holds $f(a\wedge b)=f(a)\wedge f(b)$, $f(a\vee b)=f(a)\vee f(b)$, and that the image 
of the null space and of the total space are, respectively, the null element and the identity element of the Boolean lattice.  Because each Boolean lattice can be mapped homomorphically  on a Boolean lattice of two elements, the solution to the embedding problem gives rise to the solution to the following prediction-task:  Every linear subspace of a three dimensional orthogonal vector space must  be assigned one of the values $t$ (rue), $f$ (alse) in such a way that the following requirements are met: The total space is assigned $t$, the null space $f$; if $a$ and $b$ are orthogonal subspaces then their intersection $a\wedge b$ is  assigned the value $t$ if and only if both are assigned the value $t$, and 
the subspace $a\vee b$ spanned by them is assigned the value $t$ if and only if at least one of the subspaces $a,b$ is assigned the value $t$.

An elementary geometrical argument shows that such an assignment is impossible, and that therefore it is impossible to have consistent predictions about a quantum mechanical system (not considering exceptional cases).

\end{document}